\documentclass{emulateapj}
\usepackage{natbib}
\def\s2n{S^{\prime}/N}

\shortauthors{Padoan and Nordlund}

\begin{document}
\title{The observable prestellar phase of the IMF}

\author{Paolo Padoan}
\affil{ICREA \& ICC, University of Barcelona, Marti i Franqu\`{e}s 1, E-08028 Barcelona, Spain;
ppadoan@icc.ub.edu}
\author{\AA ke Nordlund}
\affil{Centre for Star and Planet Formation and Niels Bohr Institute, University of Copenhagen,
Juliane Maries Vej 30, DK-2100, Copenhagen,
Denmark; aake@nbi.dk}

\begin{abstract}

The observed similarities between the mass function of prestellar cores (CMF) and the stellar initial mass function (IMF) 
have led to the suggestion that the IMF is already largely determined in the gas phase. However, 
theoretical arguments show that the CMF may differ significantly from the IMF.  In this Letter, we study the relation 
between the CMF and the IMF, as predicted by the IMF model of Padoan and Nordlund. We show that
1) the observed mass of prestellar cores is on average a few times smaller than that of the stellar 
systems they generate; 2) the CMF rises monotonically with decreasing mass, with a noticeable change in slope
at approximately 3-5~M$_{\odot}$, depending on mean density; 3) the selection of cores with 
masses larger than half their Bonnor-Ebert mass yields a CMF approximately consistent with the system IMF, rescaled in 
mass by the same factor as our model IMF, and therefore suitable to estimate the local efficiency of star formation, and to
study the dependence of the IMF peak on cloud properties; 4) only one in five pre-brown-dwarf core candidates is a true 
progenitor to a brown dwarf. 

\end{abstract}


\section{Introduction}

Molecular clouds (MCs) undergo a highly non-linear fragmentation process, even prior to the emergence of young stars, 
or in regions with no apparent star formation. The fundamental reason for their fragmentation is the presence of supersonic 
turbulence that originates at large scales from various sources, such as supernovae, spiral-arm shocks, or magneto-rotational 
instability. The formation of dense filaments and cores is the natural 
evolution of the intersection of randomly driven shocks in the turbulent flow. Even without self-gravity, turbulence simulations 
with the same rms Mach numbers as MCs generate density contrasts of many orders of magnitude, with characteristic size and 
density of both filaments and cores as observed in MCs \citep{Arzoumanian+11}. Because all stars are born in MCs, and 
specifically from dense cores found predominantly within the densest filaments \citep[e.g.][]{Andre+10,Arzoumanian+11}, 
the fragmentation by the turbulence must control the early phases of the star formation process. It may also control its global 
properties, such as the star formation rate \citep{Krumholz+McKee05sfr,Padoan+Nordlund11sfr} and the mass 
distribution of stars \citep{Padoan+Nordlund02imf,Hennebelle+Chabrier08}, but other processes (e.g. stellar jets and outflows, 
disk fragmentation, competitive accretion, ambipolar drift) can also affect the mass and formation rate of stars \citep[e.g.][]{Adams+Fatuzzo96,Bonnell+01comp}. 

Most (sub-)mm or extinction surveys of star-forming regions have resulted in prestellar core mass functions (CMFs) interpreted to be consistent with 
the stellar IMF \citep{Motte+98,Testi+Sargent98,Johnstone+00,Johnstone+01,Motte+01,Onishi+02,Johnstone+06,Stanke+06,Andre+07,Enoch+07,Enoch+08,Belloche+11}. A few studies have also claimed the detection of a CMF peak, which has been interpreted as a scaled progenitor of the 
stellar IMF peak, with the scaling mass factor giving the local efficiency of star formation 
\citep{Alves+07,Nutter+Ward-Thompson07,Rathborne+09,Andre+10,Konyves+10}. However, the peak of the CMF is quite close to 
the estimated completeness limits of the surveys, which depends also on the core shape and selection method 
\citep[e.g.][]{Kainulainen+09cmf,Pineda+09}. 

Simple theoretical considerations show that a one-to-one relation between the mass of observed prestellar cores and that of stars is unlikely. 
Prestellar cores must grow for some time before they can collapse and form stars. 
When a core is observed it is unlikely to have just reached its final mass. Furthermore, the largest observable mass of a core may be even 
lower than the full prestellar mass involved. Once a core mass grows beyond its Bonnor-Ebert (BE) mass \citep{Bonnor56,Ebert57}, 
the core rapidly collapses within about a free-fall time, but the accretion flow that was assembling that core is likely to still 
be active, and to continue to bring additional mass to the new protostar. The prestellar core as such is gone, but the prestellar 
mass feeding the new protostar may keep coming. Note that this process is essentially {\em inertial}, and therefore distinctly different 
from {\em competitive} accretion, which assumes the flow is caused by gravitation forces.

\cite{Clark+07imf} have stressed a timescale problem in relating the CMF to the stellar IMF: if cores of different masses are assumed to 
all contain one BE mass, small cores must be denser and free-fall more rapidly than larger ones. If the CMF is stationary in time, the 
corresponding stellar IMF should then be steeper than the CMF. However, the observations do not show a strong correlation 
between core mass and density, so this timescale issue may not be the main problem in deriving the IMF from the CMF. 
In this Letter, we study the relation between the CMF and the stellar IMF based on our IMF model \citep{Padoan+Nordlund02imf}.

\section{Random realizations of the IMF model}

The IMF model by \cite{Padoan+Nordlund02imf} (PN02 hereafter) predicts the mass distribution of gravitationally unstable cores 
generated by a turbulent flow. The mass of a core is the total mass, $m_{\rm accr}$, the turbulent flow would assemble locally, irrespective 
of when the core should collapse and cease to appear as prestellar. In PN02, the mass distribution of all cores (unstable or not) is a power 
law with slope $x=3/(4-b)$, where $b$ is the slope of the velocity power spectrum of the turbulent flow, $E(k)\sim k^{-b}$. The Salpeter 
slope, $x=1.35$, is thus recovered if $b=1.78$, consistent with power spectra derived from the largest simulations of supersonic 
and super-Alfv\'{e}nic turbulence and from observations \citep[e.g.][]{Padoan+06perseus,Padoan+09ngc1333}. Unstable cores are
then selected as those more massive than their BE mass, assuming all cores have the same temperature, while their external density 
(the local postshock density) follows the Log-Normal pdf of supersonic turbulent flows. Most of the massive cores contain more than one 
BE mass, so the IMF above approximately one solar mass is predicted to be a power law with slope close to Salpeter's. Smaller cores 
are usually less massive than their BE mass, so the IMF is expected to peak at a fraction of a solar mass, and to decline towards smaller 
masses. 

This assumes that the actual stellar mass is $m_\star=\epsilon \, m_{\rm accr}$, and the local efficiency, $\epsilon$,
is approximately independent of mass. In observational studies, the local efficiency, $\epsilon_{\rm core}$, is instead defined as the ratio 
of the resulting stellar mass, $m_\star$, and the current mass of prestellar cores, $m$, so $\epsilon_{\rm core}=m_\star / m$.
It follows that the ratio of the total core mass predicted by the model and the current core mass can be expressed as the ratio between our
theoretical local efficiency and the observational one, $m_{\rm accr}/m=\epsilon_{\rm core}/\epsilon$, and thus 
$\epsilon_{\rm core}/\epsilon \ge 1$.

To create a population of observable prestellar cores consistent with the IMF model, we first generate 
random values of total core masses,  $m_{\rm accr}$, following the predicted power law distribution with slope $x=1.35$.
We then generate a random value of external density for each core, 
according to the Log-Normal gas density distribution of the turbulent flow, converted to mass fraction,
\begin{equation}
p(\tilde{\rho})d\tilde{\rho} \propto {\rm exp} \left[ -\frac{({\rm ln}\tilde{\rho} + \sigma^2/2)^2}{2\, \sigma^2} \right] d\tilde{\rho},
\label{pdf}
\vspace{0.2cm}
\end{equation}
where $\tilde{\rho}$ is the core external density in units of the cloud mean density, $\tilde{\rho}=\rho/\rho_0$, 
and the standard deviation, $\sigma$, of the logarithmic density field is
\begin{equation}
\sigma^2= {\rm ln}\left[  1 +  \left(\frac{{\cal M}_{\rm S,0}}{2}\right)^2 (1+\beta^{-1})^{-1} \right],
\label{eq_sigma}
\end{equation}
where ${\cal M}_{\rm S,0}$ is the rms sonic Mach number, ${\cal M}_{\rm S,0} = v_0 / c_{\rm S,0}$, with $v_0$ the three-dimensional 
rms velocity and $c_{\rm S,0}$ the isothermal sound speed corresponding to the mean temperature $T_0$, and $\beta$ is a characteristic 
ratio of gas to magnetic pressure in the postshock gas \citep[see][eqs. 16, 17, 18, 27 and 28]{Padoan+Nordlund11sfr}. The 
corresponding standard deviation of the linear density field, $\sigma_{\tilde{\rho}}$, is given by
\begin{equation}
\sigma_{\tilde{\rho}}= (1+\beta^{-1})^{-1/2} {\cal M}_{\rm S,0} / 2.
\label{eq21}
\end{equation}
Finally, we associate a random age to each core, assuming for simplicity that the star formation rate (SFR) is uniform over time and 
independent of core mass. 

In order to relate the PN02 model to observations of prestellar cores (cores without a detectable embedded protostar), we also need 
to account for the time evolution of cores, and define when, during its growth, a core would be observed as prestellar. In PN02, cores 
are assumed to be chunks of dense postshock filaments (or sheets), with size equal to the postshock thickness, $\lambda$, and mass 
\begin{equation}
m_{\rm accr}=(4/3) \pi \rho ( \lambda /2)^3, 
\label{m_star}
\end{equation}
where $\rho$ is the postshock density. Assuming that a
compression from the scale $\ell$ has a shock velocity that follows the second order velocity structure function of the turbulent flow, 
$v\sim \ell^a$ (with $a=(b-1)/2$), the compression lasts for an accretion time given by 
\begin{equation}
t_{\rm accr}=\ell/v = t_0 (\ell/L_0)^{1-a},
\label{t_accr1}
\end{equation}
where $t_0$ is the cloud crossing time, defined as the ratio of cloud size, $L_0$, and rms velocity, $v_0$, $t_0=L_0/v_0$. 
Assuming that the postshock thickness, $\lambda$, grows linearly with time, we can then model the mass evolution of a prestellar 
core, up to the time $t=t_{\rm accr}$, when the total core mass, $m_{\rm accr}$, is reached, with the simple law
\begin{equation}
m(t)/ m_{\rm accr} = (t/t_{\rm accr})^3,
\label{m_core_t}
\end{equation}
which shows that cores must spend a significant fraction of their lifetime at a mass significantly lower than their final 
mass, and therefore the relation between the masses of prestellar cores and stars cannot be trivial. 

In order to express $t_{\rm accr}$  and $m$ as a function of $m_{\rm accr}$, instead of $\ell$, we can
relate $m_{\rm accr}$ and $\ell$ using equation~(\ref{m_star}) with the shock jump conditions,  $\rho=\rho_0 {\cal M}_{\rm A,\ell}$ 
and $\lambda=\ell / {\cal M}_{\rm A,\ell}$, assuming that the shock Alfv\'{e}nic Mach number scales like the shock velocity,
${\cal M}_{\rm A,\ell} \sim v \sim \ell^a$, and with the effective rms Alfv\'{e}nic Mach number expression 
of \cite{Padoan+Nordlund11sfr} (equations (13) and (14)), ${\cal M}_{\rm A,0}=\sigma_{\tilde{\rho}}^2$.
We then obtain the following expression for $t_{\rm accr}$ expressed as a function of $m_{\rm accr}$
\begin{equation}
t_{\rm accr} = t_0 \, \sigma_{\tilde{\rho}}^{{4-4a \over 3-2a}} \left( m_{\rm accr} \over m_0 \right)^{1-a \over (3-2a)}.
\label{t_accr2}
\end{equation}

We assume that cores that do not reach their BE mass are seen only during their formation time, $t_{\rm accr}$, 
while those growing past their BE mass continue to be observed as prestellar for one free-fall time, reaching a maximum
prestellar mass, $m_{\rm max}$, given by 
\begin{equation}
{m_{\rm max} \over m_{\rm accr}}=\left( {t_{\rm BE}+t_{\rm ff}} \over t_{\rm accr} \right)^3,
\label{m_max}
\end{equation}
where $t_{\rm BE}$ is the time when cores reach their BE mass,
\begin{equation}
t_{\rm BE} = t_{\rm accr} \left( m_{\rm accr} \over m_{\rm BE} \right)^{-1/3},
\label{t_BE}
\end{equation}
and the BE mass \citep{Bonnor56,Ebert57} is
\begin{equation}
m_{\rm BE} = 1.182 \, c_{\rm S}^3\, /(G^{3/2}\rho^{1/2}),
\label{m_be}
\end{equation}
where $c_{\rm S}$ is the isothermal sound speed in the cores, corresponding to the mean core temperature $T$, 
$\rho$ the postshock density (assumed to be the external density of the BE sphere), $G$ the gravitational constant,
and $t_{\rm ff}$ is the free-fall time, $t_{\rm ff}=(3 \pi/(32 G \rho))^{1/2}$. 
The value of $m_{\rm max}$ may be larger than $m_{\rm accr}$ (if $t_{\rm BE}+t_{\rm ff} > t_{\rm accr}$), and therefore the 
final (observable) mass of a prestellar core as such is
\begin{equation}
m_{\rm f} = {\rm min}[m_{\rm accr}, m_{\rm max}]. 
\label{m_f}
\end{equation}
\begin{figure}[t]
\includegraphics[width=\columnwidth]{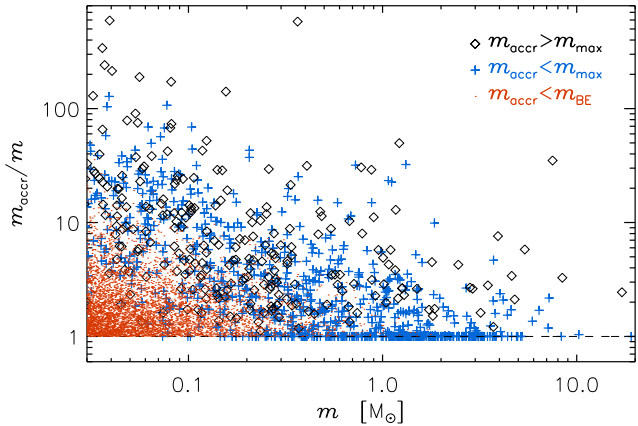}
\caption[]{Ratio of the total core mass and the current core mass, $m$. Dots are for cores that will never collapse into stars,
plus symbols for cores that form stars and can reach $m_{\rm accr}$ while prestellar, $m_{\rm f}= m_{\rm accr} < m_{\rm max}$, and
diamond symbols for those that form stars but do not reach the mass $m_{\rm accr}$ while prestellar, 
$m_{\rm f}=m_{\rm max}<m_{\rm accr}$.}
\label{mass}
\end{figure}

\section{Results}

We show results for a model with characteristic molecular cloud (MC) parameters, $T_0=10$~K, $T=7$~K (the core mean temperature), $L_0=10$~pc, 
$\rho_0=2\times10^{-21}$~g/cm$^3$, ${\cal M}_{\rm S,0}=25$, $\beta=0.4$. The total MC mass is then $m_0=1.1\times10^5$~M$_{\odot}$,
and its three-dimensional rms velocity $v_0=4.9$~km/s. We consider the simple case of a constant SFR for a time equal to the cloud crossing
time, $t_0 = L_0/v_0=1.9$~Myr, and study the prestellar core population at a time $t=t_0$, assuming that prestellar cores are detected above a 
minimum surface density of $N_{\rm det}=10^{21}$~cm$^{-2}$, corresponding to the Herschel 5-$\sigma$ detection limit due to cirrus noise 
in Aquila \citep{Andre+10,Konyves+10}. We generate a random distribution of total core masses,
$m_{\rm accr}$, with probability following a power law with slope $x=1.35$, with a minimum mass of 0.01~M$_{\odot}$. The total mass 
in the cores is $0.26 m_0$, while the total mass of those that collapse into stars gives a final  star formation efficiency of SFE$_{\rm f}$=0.05.
At the time we study the core population, $t=t_0$, the total mass in stars (defined as all the cores that have reached their total mass, $m_{\rm accr}$) 
is such that SFE$(t_0)=0.02$, which is a reasonable value for MCs \citep[e.g.][]{Evans+09}. The core formation efficiency at $t=t_0$ is 
CFE$(t_0)$=0.01, also consistent with observations \citep[e.g.][]{Enoch+07}. 

\begin{figure}[t]
\includegraphics[width=\columnwidth]{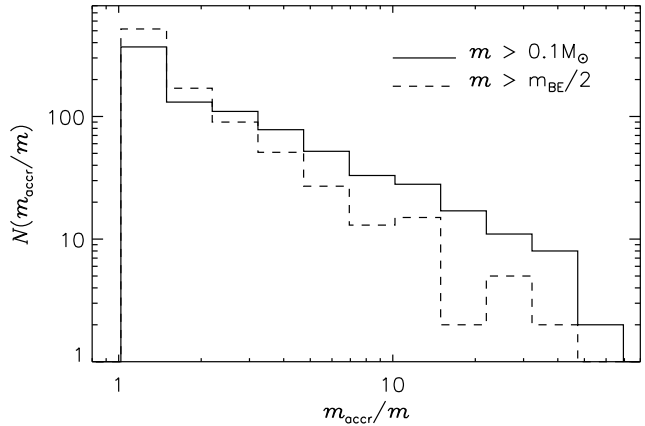}
\caption[]{Histograms of the ratio $m_{\rm accr}/m$ shown in Figure~\ref{mass}. The solid line is the histogram for 
all cores with mass $m>0.1$~M$_{\odot}$, yielding a mean value of $\langle m_{\rm accr}/m \rangle = 5.0$.
The dashed line in Figure~\ref{mstar_hist} is the histogram including only core masses above half their BE mass, 
$m > m_{\rm BE}/2$, yielding a mean value of $\langle m_{\rm accr}/m \rangle = 2.6$.}
\label{mstar_hist}
\end{figure}

Figure~\ref{mass} shows the ratio between the total core mass and its current mass at time $t=t_0$, $m_{\rm accr}/m=\epsilon_{\rm core}/\epsilon$. 
The plot shows that prestellar cores may have a mass significantly smaller than the mass of the stellar system they will form,
assuming reasonable values of $\epsilon$, which means that $\epsilon_{\rm core}>1$, contrary to the usual assumption of observational 
studies. Observed prestellar cores with masses between 0.1 and 1.0 M$_{\odot}$, for example, may be on their way to form stellar systems 
with masses 10 times larger. 

The relation between current core mass and total core mass is quantified by the histograms of the ratio 
$m_{\rm accr}/m$, shown in Figure~\ref{mstar_hist}. The solid line is the histogram for all prestellar cores with mass 
$m>0.1$~M$_{\odot}$, showing a broad distribution, with a mean value of $\langle m_{\rm accr}/m \rangle = 5.0$.
Cores that will never grow above their BE mass to form stars ($m_{\rm accr} < m_{\rm BE}$) are shown as dots in Figure~\ref{mass}. 
These cores are not included in the histogram, but they certainly contaminate observational samples. 

The dashed line in Figure~\ref{mstar_hist} is the histogram including only core masses above half their BE mass, 
$m > m_{\rm BE}/2$. These cores are somewhat closer to their corresponding total core masses, with an average ratio of 
$\langle m_{\rm accr}/m \rangle = 2.6$. The ratio of the peak of the stellar IMF from our model (see Figure~\ref{imf} below) 
and the peak of the multiple system IMF of \cite{Chabrier05} is approximately 2.1, corresponding to $\epsilon \approx 0.48$. 
With this value of the theoretical local efficiency of star formation, the average ratio between the actual mass of stellar systems, $m_\star$, 
and the current mass of cores, $m$, would be $\epsilon_{\rm core}= \epsilon \langle m_{\rm accr}/m\rangle  = 2.4$ for prestellar cores with 
$m>0.1$~M$_{\odot}$, and  $\epsilon_{\rm core} \approx 1.2$ for the core subsample with $m > m_{\rm BE}/2$.

\begin{figure}[t]
\includegraphics[width=\columnwidth]{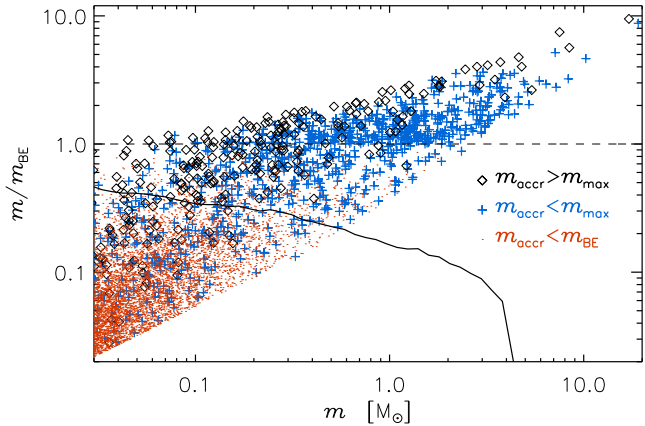}
\caption[]{Ratio of current core mass to BE mass, plotted versus the current core mass. The solid line shows the ratio between the 
number of cores that will never collapse and form stellar systems and that of true prestellar cores, after selecting cores with  
$m > m_{\rm BE}/2$.} 
\label{be_mass}
\end{figure}

Figure~\ref{be_mass} shows the ratio of the current core mass to its BE mass, at $t=t_0$. It shows that a significant
fraction of the cores with $m>0.1$~M$_{\odot}$ may be found to have a mass larger than their BE mass.  
It also shows that the selection of cores with mass $m > m_{\rm BE}/2$ allows partial decontamination 
from cores that will never collapse into stars, while still providing a large enough core subsample to allow a meaningful 
estimation of the CMF. This strategy is already being implemented in the analysis of observational surveys \citep[e.g.][]{Konyves+10}. 
The solid line in Figure~\ref{be_mass} gives the ratio between the number of non-prestellar cores and that of true prestellar cores
(computed from five different realizations of the same model), as a function of core mass, for cores with mass $m > m_{\rm BE}/2$. 
Around 1~M$_{\odot}$, only 16\% of these cores are not prestellar, while that fraction grows to 35\% around 0.1~M$_{\odot}$.

Smaller cores that may be progenitors of brown dwarfs can barely exceed their BE mass (towards the end of their lifetime). 
Only one in five of the selected pre-brown-dwarf core candidates with mass $m_{\rm BE}/2 < m < 0.15$~M$_{\odot}$ (assuming 
$\epsilon \approx 0.48$) is a true pre-brown-dwarf core. Half of them are growing progenitors 
of higher mass stars, and one third are non-prestellar cores. Distinguishing true pre-brown-dwarf cores or non-prestellar cores 
in observational surveys is a difficult task that may require the characterization of such cores using synthetic observations of turbulence 
simulations. 

The CMF at the time $t=t_0$ is shown in Figure~\ref{imf} by the black histogram (and by the continuous red line for the average of
five different realizations of the same model). The CMF seems to change its slope at two mass values, $m_1\approx 0.5$~M$_{\odot}$ 
(approximately the peak of the model IMF) and $m_2\approx 4$~M$_{\odot}$. For masses above $m_2$, the slope is steeper than $x=1.35$, 
it becomes shallower than that for masses below $m_2$, and it steepens again below $m_1$, but still remaining a bit shallower than $x=1.35$. 
The slope for masses below $m_1$ is presumably very difficult to derive from observations, because core samples are generally incomplete 
at such low masses, depending on the survey sensitivity, confusion, and method of core selection. 

We have verified that the values of $m_1$
and $m_2$ scale approximately as $\rho_0^{1/2}$ ($m_1$ remains close to the peak of the model IMF), while they are approximately the same 
in models where equation (\ref{m_core_t}) is modified to assume $m(t) \sim t^2$, and in models where cores are assumed to remain prestellar 
for only a fraction $f$ of $t_{\rm ff}$ after they reach their BE mass. However, with decreasing values of $f$, the slope of the CMF for masses 
in the range $m_1< m < m_2$ increases, becoming approximately the same as the slope for $m < m_1$ when $f\approx 0.6$, and approximately 
equal to 1.35 when $f \le 0.4$. 

The model IMF seen at the time $t=2t_0$, when all cores have reached their total mass, $m_{\rm accr}$, and have already collapsed into stars 
(with SFE$_{\rm f}=0.05$), is shown by the green histogram. This model IMF is consistent with the system IMF of \cite{Chabrier05}, with 
the masses shifted by a factor of 2.1, shown in Figure~\ref{imf} as a dashed line (connected to a Salpeter IMF at 2~M$_{\odot}$). This shift in mass 
corresponds to $\epsilon=0.48$.

Figure~\ref{imf} shows that the observed CMF should rise monotonically towards smaller masses, even past the peak of the stellar IMF.
However, a careful comparison 
between the model and the observations requires a detailed consideration of sensitivity, uncertainties, and method of core selection of each 
survey, besides the use of specific physical parameters and age for the model, matching those of the observed regions. Furthermore, a calculation of 
the completeness of the surveys may benefit from a model-dependent assumption on the mass and density distributions of small cores 
(near the detection limit), for example to estimate the confusion arising from projection effects. Such studies are beyond the scope of 
this Letter and should be addressed in future works. 

As mentioned in the discussion of Figure~\ref{mstar_hist}, prestellar core masses would be statistically better related to their corresponding
stellar system masses if one were able to select only cores with mass $m > m_{\rm BE}/2$. This is illustrated in Figure~\ref{imf}
by the CMF of such core subsample (blue histogram), which is fit by a \cite{Chabrier05} system IMF with masses shifted by 
a factor of 2.1 (dashed line), except for the largest masses, where the slope is steeper than Salpeter's. This CMF peaks at the same mass as the
model system IMF, and thus may be used to estimate the value of $\epsilon$ (because in this case $\epsilon_{\rm core} \approx \epsilon$, based
on the peaks alone). Its peak could also be used to study possible variations of $\epsilon$ with cloud properties. \\

\begin{figure}[t]
\includegraphics[width=\columnwidth]{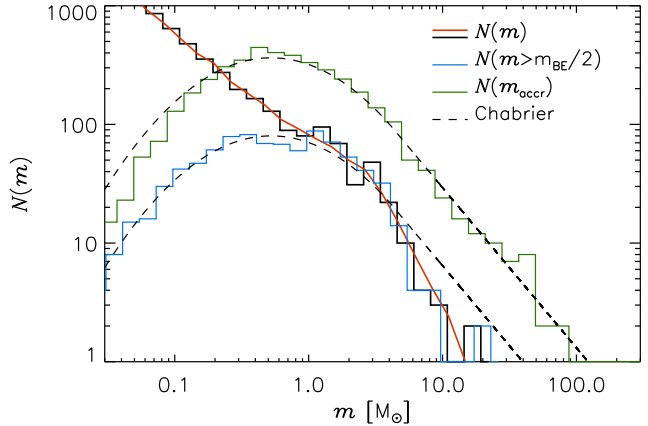}
\caption[]{Black histogram: CMF at time $t=t_0$. Blue histogram: CMF for a prestellar core subsample with masses $m(t_0)> m_{\rm BE}/2$. 
The same CMF derived from five different realizations of the same model is shown by the red solid line. Green histogram: Model IMF at the 
time $t=2 t_0$. Dashed lines: \cite{Chabrier05} system IMFs, shifted in mass by a factor of 2.1.}
\label{imf}
\end{figure}

\section{Conclusions}

We have computed the observable CMF predicted by the PN02 IMF model, assuming characteristic MC parameters
that are shown to yield a \cite{Chabrier05} system IMF with $\epsilon \approx 0.48$. 
Our main results are the following: 1) The observed mass of prestellar cores is on average a few times smaller than that of the stellar 
systems they generate, so $\epsilon_{\rm core}>1$. 2) The CMF rises monotonically with decreasing 
mass, and shows a noticeable change in slope at approximately 3-5~M$_{\odot}$, depending on mean density; 3) the selection of cores with 
masses $m > m_{\rm BE}/2$ yields a CMF approximately consistent with the system IMF, rescaled in mass by the same factor as the model IMF, 
and therefore suitable to estimate the local efficiency of star formation, $\epsilon$, and its possible dependence on cloud properties; 
4) Only one in five pre-brown-dwarf core candidates is a true progenitor to a brown dwarf. 

We have not discussed the {\it proto}stellar CMF, nor the system IMF at time $t=t_0$ (when relatively massive protostars are still growing 
in mass). The definition of the protostellar phase faces the difficulty of accounting for the rate of mass transfer between the core and the 
protostar \citep{McKee+Offner10}. However, we have verified that with a loose definition of protostars as all the cores 
that have reached their final prestellar mass, $m_{\rm max}$, but not yet their total mass, $m_{\rm accr}$ (core ages between 
$t_{\rm BE}+t_{\rm ff}$ and  $t_{\rm accr}$), the protostellar CMF is significantly shallower than the prestellar CMF, in agreement 
with the observations \citep[e.g.][]{Hatchell+Fuller08,Enoch+08}, and the current system IMF has a high-mass slope a bit steeper 
than Salpeter's (not observationally tested yet, to our knowledge), due to the fact that more massive cores remain longer in the protostellar 
phase than lower mass ones.

Upcoming starless core samples from the Herschel Gould belt survey \citep{Andre+10} will allow comparisons with our model 
predictions. In this Letter we have only discussed the results of a very large core sample and a specific set of cloud parameters, without 
simulating observational uncertainties and incompleteness, and thus setting aside a detailed comparison with observed CMFs for a separate work.

\acknowledgements

We thank Alyssa Goodman for reading the manuscript and providing comments, and the anonymous referee for useful comments and corrections. 
PP is supported by the Spanish MICINN grant AYA2010-16833 and by the FP7-PEOPLE-2010-RG grant PIRG07-GA-2010-261359. 
The work of {\AA}N is supported by the Danish National Research Foundation, through its establishment of the Centre for Star and Planet Formation.

\bibliographystyle{apj}

\end{document}